\documentclass[Aps,prl,reprint,twocolumn,showpacs,groupedaddress]{revtex4-2}
\usepackage{amssymb}
\usepackage[utf8]{inputenc}
\usepackage{float}
\usepackage{color}
\usepackage{soul}
\usepackage[pdftex,colorlinks=true,linkcolor=red,citecolor=blue]{hyperref}

\usepackage{graphicx}
\usepackage{dcolumn}
\usepackage{bm}

\begin{document}
\title[]{Tuning of electronic properties in highly lattice-mismatched epitaxial SmN}%

\author{Kevin D. Vallejo$^{1,2}$}
\email{kevin.vallejo@inl.gov}
\author{Volodymyr Buturlim$^{3}$}
\author{Zach E. Cresswell$^{1}$}
\author{Brooke Campbell$^{1}$}
\author{Bobby G. Duersch$^{4}$}
\author{Brelon J. May$^{1}$}
\author{Krzysztof Gofryk$^{2,3}$}
\email{gofryk@inl.gov}

\affiliation{$^{1}$Idaho National Laboratory, Idaho Falls, Idaho 83415, USA}
\affiliation{$^{2}$Center for Quantum Actinide Science and Technology, Idaho National Laboratory, Idaho Falls, Idaho 83415, USA}
\affiliation{$^{3}$Glenn T. Seaborg Institute, Idaho National Laboratory, Idaho Falls, Idaho 83415, USA}
\affiliation{$^{4}$Utah Nanofab Electron Microscopy and Surface Analysis Laboratory, University of Utah, Salt Lake City, Utah 84112, USA}

\date{\today}

\begin{abstract}
We demonstrate that the electronic properties of epitaxial SmN thin films can be effectively tuned during growth by controlling the synthesis parameters. By carefully adjusting these parameters, we are able to drive SmN from an insulating ferromagnetic state to a ferromagnetic metallic state. However, no signatures of previously reported superconductivity were observed down to 0.35 K, even in the most conductive samples. We discuss possible scenarios for the absence of superconductivity in these films and examine implications for the underlying pairing mechanism in this material. These findings open a new pathway for the epitaxial engineering of multifunctional materials, enabling the monolithic integration of diverse electronic phases, such as ferromagnetism and metallicity, without the lattice mismatch and strain typically associated with heteroepitaxial growth of dissimilar materials.
\end{abstract}

\maketitle


\textit{Introduction} - Rare-earth nitrides (RENs) exhibit a wide variety of properties desirable in the field of spintronics, infrared detectors, intrinsically ferromagnetic-based tunnel junctions, and as strongly correlated electron materials \cite{Natali2013}. The electronic configuration of elements containing 4$f$ orbitals is a source of interesting new physics: as an example, samarium nitride (SmN) has been reported to support the coexistence of semiconductor behavior, ferromagnetic states, and unconventional superconductivity \cite{Anton2016}. Motivated by these properties and exciting opportunities, presently there is an increase in interest on the synthesis and study of high-quality REN materials \cite{moon1979magnetic,preston2007comparison,Anton2016,Holmes2018,Holmes2019,Azeem2018,Azeem2019,Natali2012,Anton2013, McNulty2015,Holmes2018, McNulty2021,Chan2016,Vezian2016,Anton2023,melendez2024influence,vallejo2024synthesis}. 

Ferromagnetic semiconductors possess magnetic properties ideal for enabling spintronic device architectures \cite{xie2025emerging}. These devices aim to leverage the electrons' spin, in addition to their charge, to reduce the energy consumption of memory and logic devices. To this end, to effectively polarize, inject, store, manipulate, and read magnetic and spin information requires devices with low-defect densities and pristine interfaces \cite{akinaga2002semiconductor,dietl2007ferromagnetic}. Furthermore, ferromagnetic semiconductors have been shown to exhibit unique phenomena such as the anomalous Hall effect, spin current polarization, and magneto-optical effects, which are predicted to bypass limitations currently facing conventional electronic device architectures. 
The spin-orbit coupling in the Sm ions is evident in the orbital-dominated nature of the ferromagnetism in SmN, which makes it an ideal candidate for spintronic devices, demonstrated in several designs with GdN \cite{pot2024magnetic,McNulty2015}. Extending the architecture of ferromagnetic and/or superconducting devices (giant magnetoresistance devices and Josephson junctions) to include antiferromagnetic materials has also been of recent interest in the field of spintronics \cite{shao2024antiferromagnetic,xie2025emerging}. Identifying growth conditions that allow tuning of its electronic properties often allows to narrow down the search for growth conditions of similar systems. Such is the case of SmN, recently realized on a largely mismatched substrate that enabled the highest structural quality film reported to date \cite{vallejo2024synthesis}. The potential coexistence of ferromagnetism and triplet superconductivity within spin-polarized bands in this compound, demonstrated to be able to be synthesized in lattice parameters compatible with transition metal nitride compounds, can enable new physics and applications. 

In this study we show how the electronic properties of SmN$_x$ (henceforth SmN) thin films grown on MgO(001) via molecular beam epitaxy (MBE) by tuning metal to nitrogen ratio and substrate temperature to create different electronic behaviors attributed to different levels of carrier concentration. We find that substrate temperature (T$_{sub}$) impacts the availability of carriers in a much more pronounced way than N flow. Furthermore, by changing the growth conditions we can tune SmN from ferromagnetic insulating to ferromagnetic metallic states. The ferromagnetism is clearly observed for all our samples. However, we did not observe any sign of superconductivity down to 0.35 K, even for the most conductive samples. This observation reopens discussions on the possible nature of superconductivity and the pairing mechanism in this material. These results indicate a path forward in the epitaxy of versatile materials able to provide monolithic integration of different electronic behaviors without the associated strain brought about by heteroepitaxial growth of dissimilar materials.

\begin{figure*}
\centering
\includegraphics[width=\textwidth]{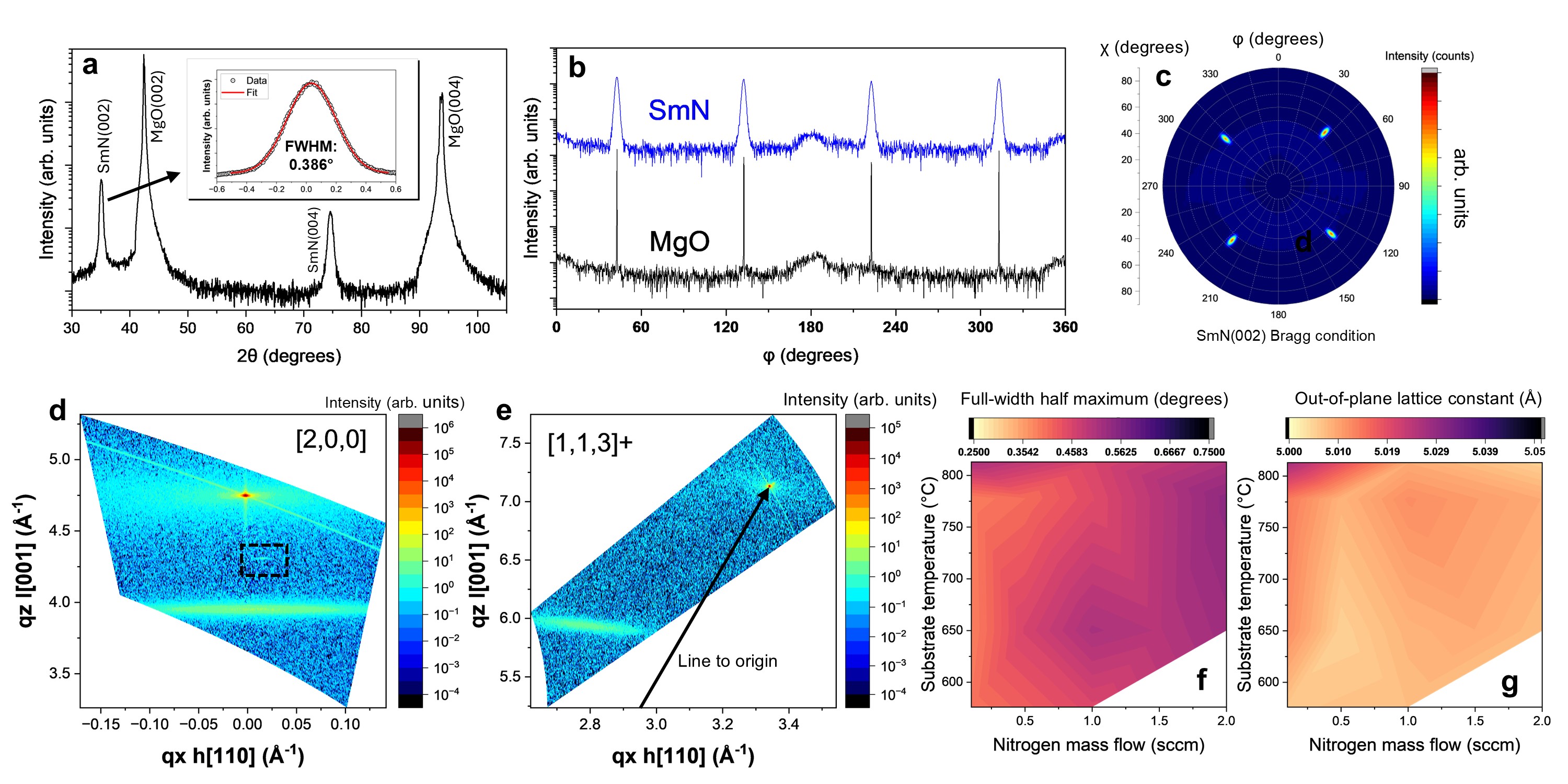}
\caption{Structural characterization of a SmN sample with smallest FWHM in the series. (a) 2$\theta$-$\omega$ survey scan showcasing the absence of any film peaks beyond SmN and substrate, with an inset showing FWHM calculation of SmN(002) peak. (b) Phi scan showing the registry between film and substrate, showing negligible strain between layers. (c) Pole figure at SmN(002) Bragg conditions for the SmN layer. (d) Symmetric reciprocal space map along [2,0,0]. The black dashed area shows the presence of a highly strained layer, possibly some Sm-O-N alloy, between the substrate and SmN film. (e) Asymmetric reciprocal space map along [1,1,3]+. (f) FWHM as a function of growth parameters for various SmN samples grown with varying N mass flows and T$_{sub}$. Out-of-plane lattice constant as a function of growth parameters.\label{fig_1}}
\end{figure*}


\textit{Methods} - In the present study we identified synthesis conditions that preserved structural quality of SmN while varying the number of available charge carriers and observe their influence on the electrical and magnetic properties of the films. To this end we synthesized all samples using variations of the procedures reported previously \cite{vallejo2024synthesis}. Substrate temperatures (T$_{sub}$) range from 576-813°C, and nitrogen was supplied using a RF plasma source with a varying mass flow of 0.1 - 2.0 standard cubic centimeter (sccm) and 250 W of power. After the completion of the 30 nm SmN layer, samples were capped with 3 nm of CrN grown at 650°C, and then cooled down to 300°C under N overpressure. We systematically varied the substrate temperature and available N flux. \textit{In-situ} structural characterization included reflection high-energy electron diffraction (RHEED), while \textit{ex-situ} structural X-ray diffraction characterization employed Cu K$\alpha_1$ radiation ($\lambda$ = 1.5406 \AA). A contouring Deaulaney triangulation algorithm generates the interpolations used in the T$_{sub}$ versus nitrogen mass flow maps, created in the OriginPro software package. The electrical properties were first characterized by room-temperature transport carried out in a van der Pauw configuration on a Nanomagnetics ezHEMS (Hall Effect Measurement System). Electrical resistivity at low temperatures was measured using a Quantum Design DynaCool-14 measurement System (PPMS) using the 4-probe technique. The resistivity was measured in the temperature range of 300 – 2 K and maximum magnetic field of 12 T. For selected samples, a PPMS He3 option was used for the measurement of the magnetotransport properties down to 0.35 K. Sample quality, composition, stability were measured using XPS analysis with a monochromated Al K$\alpha$ X-ray source. 

\begin{figure*}
\centering
\includegraphics[width=\textwidth]{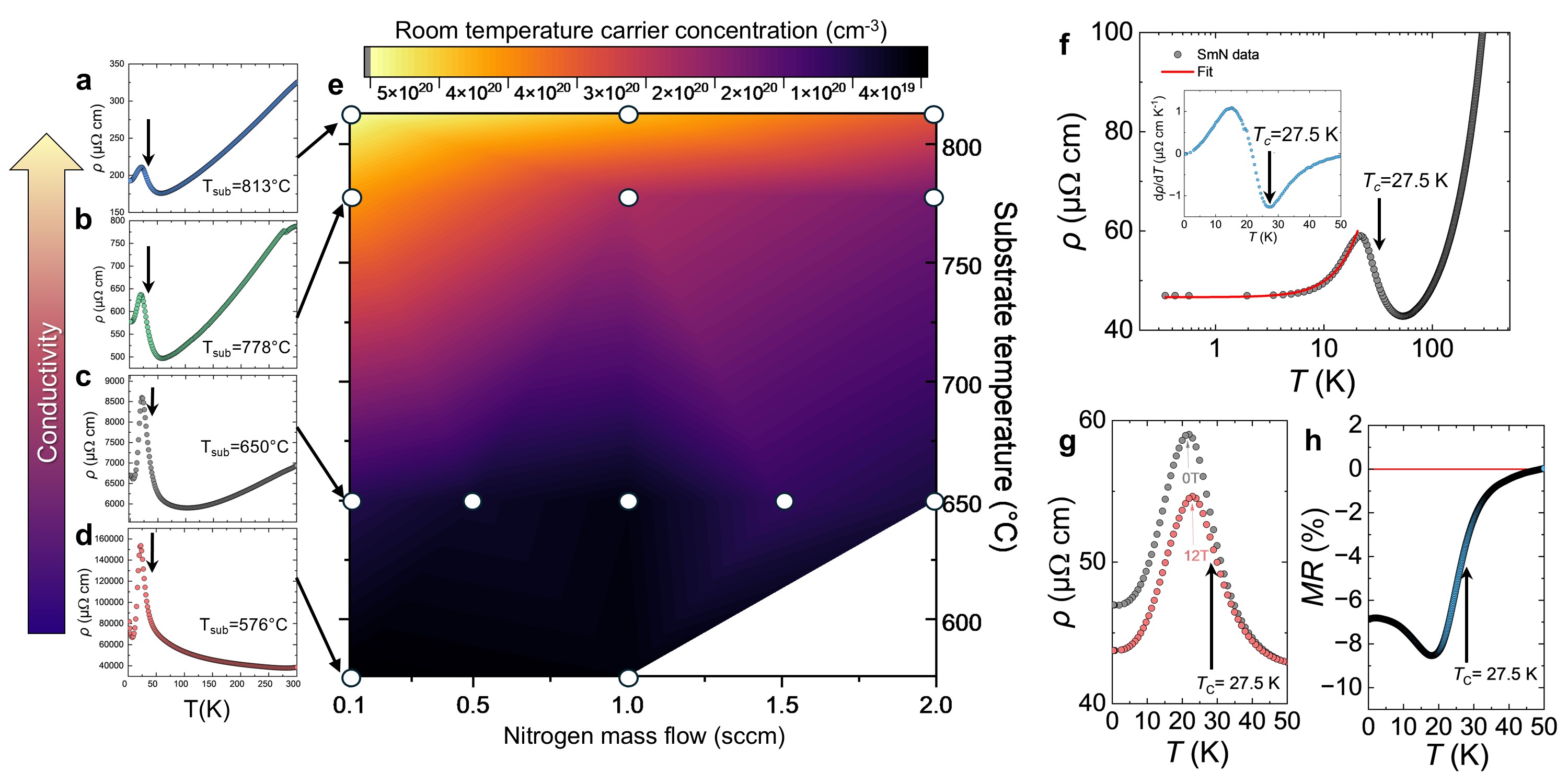}
\caption{(a)-(d) Resistivity vs. temperature plots of SmN samples grown at different substrate temperatures with every other growth parameter kept identical. Lower substrate temperature results in lower carrier concentrations and more resistive samples. Arrows indicate Curie temperature. (e) Carrier concentration as a function of nitrogen mass flow and substrate temperature measured at room temperature. Fourteen samples (indicated with white circles) are used to generate this map. (f) Resistivity as a function of temperature for a SmN sample grown using a GaN capping layer. A ferromagnetic ordering is observed below 27.5 K, determined as a minimum in using the first derivative with respect to temperature (inset). (g) Low temperature dependence of the electrical resistivity measured in 0 and 12 T magnetic fields. (h) Magnetoresistivity vs. temperature obtained for H = 12 T. The arrow shows a ferromagnetic phase transition. \label{fig_2}}
\end{figure*}

\textit{Results and Discussion} - Our results show that SmN thin films can be grown on the highly mismatched MgO(001) substrate, and their electronic properties can be tuned without affecting the structural quality of the sample. Long-range 2$\theta - \omega$ X-ray diffraction scans found no evidence of other phases besides SmN and MgO (Fig.~\ref{fig_1}(a)), with an average SmN(002) peak full-width half-maximum (FWHM) of 0.46°$\pm$0.12°, and the best sample having a 0.386°$\pm$0.001° FWHM (inset to Fig.~\ref{fig_1}(a)). All samples present a comparable degree of crystallinity and structural quality evidenced by the values of FWHM and SmN(002) peak position (Fig.~\ref{fig_1}(b) and (c)). Despite our synthesis being designed to introduce N vacancies, all samples preserved a highly crystalline SmN rock salt structure. We further confirmed the SmN films to present no texture despite the potential presence of vacancies, as evidenced by Fig.~\ref{fig_1}(d). 

The symmetric reciprocal space map (RSM) taken in the [2,0,0] direction shows the highest intensity peak at q$_z$ = 4.75 \AA$^{-1}$ (MgO substrate) and a broad peak at q$_z$ = 3.95 \AA$^{-1}$ corresponding to SmN. From these peak positions, the calculated lattice constant presents negligible differences (Fig.~\ref{fig_1}(g).) The seemingly unstrained SmN can be explained by the presence of a third peak in nearly every sample in the series (highlighted in dashed black rectangle). This peak occurs at q$_z$ = 4.64 \AA$^{-1}$ (roughly 38.8°). Its position outside of the line between MgO and SmN suggest a Sm-N-O compound forming at the interface, or a highly compressively strained Sm-O layer \cite{felmlee1968ternary}. The position  could be attributed to the [001] planes of this layer being tilted relative to the [001] planes of SmN and MgO, explaining their absence in the normal incidence wide angle scans and the dimming of the RHEED diffraction pattern during the initial stages of growth \cite{vallejo2024synthesis}. The asymmetric RSM taken in the [1,1,3]+ direction (Fig.~\ref{fig_1}(f)) shows only MgO and SmN peaks, whose positions confirm the strain-free state of the SmN film. Phi scans of the samples taken for substrate and film show peaks at 90° intervals and in alignment, as seen in Fig.~\ref{fig_1}(g), confirming the epitaxial relation between film and substrate. While the nature of the strain-relieving mechanism is still under debate, it is important to highlight this material's small FWHM at such different growth conditions, and the fact that MgO is a platform on which transition metal nitride systems can be grown epitaxially makes this integration quite promising. The combined formation of the intermediate layer and potentially an array of defects (e.g., interfacial misfit dislocation arrays \cite{nordstrom2023direct,zhang2024growth}) could be the mechanism responsible for the relief of strain. 

Vacancy quantification using X-ray photoelectron spectroscopy showed a large chromium oxide layer as well as Mg and Sm signatures intermixed with the Cr signal, which indicated a sample degradation by the time these measurements were performed.  These results allowed us to identify the degradation and oxidation processes of SmN. The electrical characterization confirmed the tunability in electrical properties of the samples. We measured resistivity as a function of temperature ($\rho$(T)) of the samples with the highest expected charge carrier density, i.e. those grown with the least amount of N mass flow (0.1 sccm). The temperature dependence of $\rho$(T) synthesized at T$_{sub}$ of 576°C, 650°C, 778°C and 813°C are shown on Fig.~\ref{fig_2} (a)-(d). The sample grown at 576°C exhibits a negative temperature coefficient of resistivity, characteristic of semiconducting materials. On the other hand, SmN films grown at higher temperatures show metallic (positive temperature coefficient of resistivity) behavior above 50 K, which indicate that the films become degenerately doped semiconductors, with decreasing resistivity as T$_{sub}$ increases. All samples exhibit a pronounced peak in $\rho$(T), associated with the transition to a ferromagnetic state, governed by the spin and orbital moments of the ground state of the Sm$^{3+}$ ion \cite{Anton2013}. The transition temperature was determined from the peak in derivative d$\rho$/d\textit{T} located at roughly 27.5 K (see inset of the Fig.~\ref{fig_2} (f)), in agreements with previous estimates of the Curie temperature (27 - 37 K) in this material\cite{McNulty2021,Meyer2008}. For semiconducting samples, the energy gaps estimated in the paramagnetic state are relatively small, being in the range 4 - 8  meV. At the highest temperature (813°C) and lowest available N mass flow (0.1 sccm) we observe a residual resistivity of 190 $\mu \Omega { cm}$, and a RRR $\rho_{300 K}$ / $\rho_{2 K}$ = 1.69. 

Fig.~\ref{fig_2}(e) shows the carrier density measured as function of T$_{sub}$ and N mass flow. Earlier reports on the SmN system indicate carrier density as the main, or possibly only, parameter which influences transport properties and magnetism of the material. Theoretical calculations suggest how the 6 Sm atoms that coordinate N$_V$ host roughly one of three released electrons, and these Sm atoms exhibit 4$f$ states which fall to the Fermi energy, subsequently hybridizing with Sm 5$d$ and N 2$p$ states and affecting electron transport \cite{Holmes2023}. Higher T$_{sub}$ is reported to decrease the N absorption on the crystal surface (\cite{karpov2000surface}) while a decrease in N mass flow limits the number of atoms available for incorporation, compounding the increase of N$_V$. Given that the Sm-limited growth rate is not affected, this increased number in N$_V$ enable carriers to be available for electrical conduction: we observe this trend indirectly through an increase in carrier density from 4.14$\times$10$^{19}$ cm$^{-3}$ to 5.51$\times$10$^{20}$ cm$^{-3}$ as T$_{sub}$ increases from 576° to 813°. This increase in charge carriers consequently influences electron transport properties, allowing the SmN layer to transition from slightly insulating to metallic-like behavior \cite{martin2017nanotechnology}. 

Below the magnetic ordering the electrical resistivity of the film with the highest carrier concentration of 5.51$\times$10$^{20}$ cm$^{-3}$ (grown at T$_{sub}$=813°C and 0.1 sccm) can be described by the expression: $\rho(T) = \rho_0 + AT^2$, characteristic of ferromagnetic systems \cite{gupta2012magnetocaloric}, with the $\rho_0$ = 46.7 $\mu \Omega { cm}$ and \textit{A} = 3.2 $\times$  $10^{-8}$ $\Omega { cm}$  $K^{-2}$).  The presence of the ferromagnetic ordering is further supported by the negative magnetoresistivity observed below $T_{C}$, as shown in Figs.~\ref{fig_2}(g-h). As can be seen from Figs.~\ref{fig_2}(f), no sign of a superconducting state is observed down to the lowest temperature measured, 0.35 K. This absence of superconductivity could be related to carrier concentration and/or inhomogeneity. Superconductivity observed in polycrystalline SmN might be sensitive to defect-induced doping (grain boundaries, point defects, and anti-site domains), which may be less tunable or more uniform in single-crystalline thin films. Polycrystalline samples may also contain inhomogeneous regions that might percolate superconductivity and this mechanism doesn’t easily translate to high-quality thin films. It is worth mentioning that our carrier concentration is an order of magnitude lower than that of reported superconductive samples with similar properties \cite{McNulty2021}, n$_i$ = 2$\times$10$^{21}$ cm$^{-3}$. Carrier mobility is another parameter that correlates with the availability of free carriers, having found a maximum value of 52.68 cm$^2$/V-s at T$_{sub}$ = 813°C, and a low value of 25.4 cm$^2$/V-s at T$_{sub}$ = 576°C. Furthermore, thin films experience substrate-induced strain, which likely alters the subtle balance between the spin-split conduction band and the chemical potential. This balance is critical for triplet pairing to emerge in a half-metallic systems. In addition to this, the spin-triplet will be highly susceptible to disorder and interface perturbations and even minimal surface or interface scattering in thin films can suppress the pairing. Finally, in thin films, quantum confinement can alter the band structure, pushing the Fermi level out of the optimal regime for pairing, preventing superconductivity from emerging. 


\textit{Summary and Conclusions} - We demonstrate the ability to tune the electronic transport properties of SmN grown on highly mismatched MgO(001) via molecular beam epitaxy over three orders of magnitude. By varying growth conditions, we observe N incorporation is much more susceptible to T$_{sub}$ than to the available N mass flow. All samples show a very pronounced anomaly at the ferromagnetic transition in the electrical resistivity. The presence of the ferromagnetic state is further supported by a characteristic temperature dependence of the electrical resistivity and a negative magnetoresistivity below the Curie temperature. In all our samples we did not observed any sign of superconductivity \cite{Anton2016} and we discuss various causes and their relationship to the nature of superconductivity in these materials and potential pairing mechanisms. Furthermore, the tunability and high structural quality of SmN makes it a promising material system for magnetism, superconductivity, and spintronics applications.

\textit{Acknowledgments} - This work was supported through the INL Laboratory Directed Research \& Development Program under U.S. Department of Energy Idaho Operations Office Contract DE-AC07-05ID14517. ZEC acknowledges the support of the NNSA Minority Serving Institution Partnership Program (MSIPP). This work made use of Nanofab EMSAL shared facilities of the Micron Technology Foundation Inc. Microscopy Suite sponsored by the John and Marcia Price College of Engineering, Health Sciences Center, Office of the Vice President for Research. Acquisition of the Bruker D8 Discover system was made possible by the Air Force Office of Scientific Research under project number FA9550-21-1-0293

\textit{Author contributions} - Conceptualization, K.V. and K.G.; methodology, K.V., K.G., and B.M.; formal analysis, K.V., B.M., K.G., V.B., B.G., B.C., and Z.C.; investigation, K.V. and V.B.; resources, K.V. and K.G.; data curation, K.V.; writing---original draft preparation, K.V.; writing---review and editing, K.V., B.M., K.G., V.B., and Z.C.; project administration, K.V.; funding acquisition, K.V. All authors have read and agreed to the published version of the manuscript.

\textit{Data Availability} - The raw data supporting the conclusions of this article will be made available by the authors on reasonable request.

\textit{Conflict of interests -} The authors declare no conflicts of interest.

\bibliography{main}

\providecommand{\noopsort}[1]{}\providecommand{\singleletter}[1]{#1}%
\begin{thebibliography}{30}%
\makeatletter
\providecommand \@ifxundefined [1]{%
 \@ifx{#1\undefined}
}%
\providecommand \@ifnum [1]{%
 \ifnum #1\expandafter \@firstoftwo
 \else \expandafter \@secondoftwo
 \fi
}%
\providecommand \@ifx [1]{%
 \ifx #1\expandafter \@firstoftwo
 \else \expandafter \@secondoftwo
 \fi
}%
\providecommand \natexlab [1]{#1}%
\providecommand \enquote  [1]{``#1''}%
\providecommand \bibnamefont  [1]{#1}%
\providecommand \bibfnamefont [1]{#1}%
\providecommand \citenamefont [1]{#1}%
\providecommand \href@noop [0]{\@secondoftwo}%
\providecommand \href [0]{\begingroup \@sanitize@url \@href}%
\providecommand \@href[1]{\@@startlink{#1}\@@href}%
\providecommand \@@href[1]{\endgroup#1\@@endlink}%
\providecommand \@sanitize@url [0]{\catcode `\\12\catcode `\$12\catcode `\&12\catcode `\#12\catcode `\^12\catcode `\_12\catcode `\%12\relax}%
\providecommand \@@startlink[1]{}%
\providecommand \@@endlink[0]{}%
\providecommand \url  [0]{\begingroup\@sanitize@url \@url }%
\providecommand \@url [1]{\endgroup\@href {#1}{\urlprefix }}%
\providecommand \urlprefix  [0]{URL }%
\providecommand \Eprint [0]{\href }%
\providecommand \doibase [0]{https://doi.org/}%
\providecommand \selectlanguage [0]{\@gobble}%
\providecommand \bibinfo  [0]{\@secondoftwo}%
\providecommand \bibfield  [0]{\@secondoftwo}%
\providecommand \translation [1]{[#1]}%
\providecommand \BibitemOpen [0]{}%
\providecommand \bibitemStop [0]{}%
\providecommand \bibitemNoStop [0]{.\EOS\space}%
\providecommand \EOS [0]{\spacefactor3000\relax}%
\providecommand \BibitemShut  [1]{\csname bibitem#1\endcsname}%
\let\auto@bib@innerbib\@empty
\bibitem [{\citenamefont {Natali}\ \emph {et~al.}(2013)\citenamefont {Natali}, \citenamefont {Ruck}, \citenamefont {Plank}, \citenamefont {Trodahl}, \citenamefont {Granville}, \citenamefont {Meyer},\ and\ \citenamefont {Lambrecht}}]{Natali2013}%
  \BibitemOpen
  \bibfield  {author} {\bibinfo {author} {\bibfnamefont {F.}~\bibnamefont {Natali}}, \bibinfo {author} {\bibfnamefont {B.~J.}\ \bibnamefont {Ruck}}, \bibinfo {author} {\bibfnamefont {N.~O.}\ \bibnamefont {Plank}}, \bibinfo {author} {\bibfnamefont {H.~J.}\ \bibnamefont {Trodahl}}, \bibinfo {author} {\bibfnamefont {S.}~\bibnamefont {Granville}}, \bibinfo {author} {\bibfnamefont {C.}~\bibnamefont {Meyer}},\ and\ \bibinfo {author} {\bibfnamefont {W.~R.}\ \bibnamefont {Lambrecht}},\ }\href {https://doi.org/10.1016/j.pmatsci.2013.06.002} {\bibinfo {title} {Rare-earth mononitrides}} (\bibinfo {year} {2013})\BibitemShut {NoStop}%
\bibitem [{\citenamefont {Anton}\ \emph {et~al.}(2016)\citenamefont {Anton}, \citenamefont {Granville}, \citenamefont {Engel}, \citenamefont {Chong}, \citenamefont {Governale}, \citenamefont {Zülicke}, \citenamefont {Moghaddam}, \citenamefont {Trodahl}, \citenamefont {Natali}, \citenamefont {Vézian},\ and\ \citenamefont {Ruck}}]{Anton2016}%
  \BibitemOpen
  \bibfield  {author} {\bibinfo {author} {\bibfnamefont {E.~M.}\ \bibnamefont {Anton}}, \bibinfo {author} {\bibfnamefont {S.}~\bibnamefont {Granville}}, \bibinfo {author} {\bibfnamefont {A.}~\bibnamefont {Engel}}, \bibinfo {author} {\bibfnamefont {S.~V.}\ \bibnamefont {Chong}}, \bibinfo {author} {\bibfnamefont {M.}~\bibnamefont {Governale}}, \bibinfo {author} {\bibfnamefont {U.}~\bibnamefont {Zülicke}}, \bibinfo {author} {\bibfnamefont {A.~G.}\ \bibnamefont {Moghaddam}}, \bibinfo {author} {\bibfnamefont {H.~J.}\ \bibnamefont {Trodahl}}, \bibinfo {author} {\bibfnamefont {F.}~\bibnamefont {Natali}}, \bibinfo {author} {\bibfnamefont {S.}~\bibnamefont {Vézian}},\ and\ \bibinfo {author} {\bibfnamefont {B.~J.}\ \bibnamefont {Ruck}},\ }\bibfield  {title} {\bibinfo {title} {Superconductivity in the ferromagnetic semiconductor samarium nitride},\ }\bibfield  {journal} {\bibinfo  {journal} {Physical Review B}\ }\textbf {\bibinfo {volume} {94}},\ \href {https://doi.org/10.1103/PhysRevB.94.024106}
  {10.1103/PhysRevB.94.024106} (\bibinfo {year} {2016})\BibitemShut {NoStop}%
\bibitem [{\citenamefont {Moon}\ and\ \citenamefont {Koehler}(1979)}]{moon1979magnetic}%
  \BibitemOpen
  \bibfield  {author} {\bibinfo {author} {\bibfnamefont {R.}~\bibnamefont {Moon}}\ and\ \bibinfo {author} {\bibfnamefont {W.}~\bibnamefont {Koehler}},\ }\bibfield  {title} {\bibinfo {title} {Magnetic properties of {SmN}},\ }\href@noop {} {\bibfield  {journal} {\bibinfo  {journal} {Journal of Magnetism and Magnetic Materials}\ }\textbf {\bibinfo {volume} {14}},\ \bibinfo {pages} {265} (\bibinfo {year} {1979})}\BibitemShut {NoStop}%
\bibitem [{\citenamefont {Preston}\ \emph {et~al.}(2007)\citenamefont {Preston}, \citenamefont {Granville}, \citenamefont {Housden}, \citenamefont {Ludbrook}, \citenamefont {Ruck}, \citenamefont {Trodahl}, \citenamefont {Bittar}, \citenamefont {Williams}, \citenamefont {Downes}, \citenamefont {DeMasi} \emph {et~al.}}]{preston2007comparison}%
  \BibitemOpen
  \bibfield  {author} {\bibinfo {author} {\bibfnamefont {A.}~\bibnamefont {Preston}}, \bibinfo {author} {\bibfnamefont {S.}~\bibnamefont {Granville}}, \bibinfo {author} {\bibfnamefont {D.}~\bibnamefont {Housden}}, \bibinfo {author} {\bibfnamefont {B.}~\bibnamefont {Ludbrook}}, \bibinfo {author} {\bibfnamefont {B.}~\bibnamefont {Ruck}}, \bibinfo {author} {\bibfnamefont {H.}~\bibnamefont {Trodahl}}, \bibinfo {author} {\bibfnamefont {A.}~\bibnamefont {Bittar}}, \bibinfo {author} {\bibfnamefont {G.}~\bibnamefont {Williams}}, \bibinfo {author} {\bibfnamefont {J.}~\bibnamefont {Downes}}, \bibinfo {author} {\bibfnamefont {A.}~\bibnamefont {DeMasi}}, \emph {et~al.},\ }\bibfield  {title} {\bibinfo {title} {Comparison between experiment and calculated band structures for {DyN} and {SmN}},\ }\href@noop {} {\bibfield  {journal} {\bibinfo  {journal} {Physical Review B—Condensed Matter and Materials Physics}\ }\textbf {\bibinfo {volume} {76}},\ \bibinfo {pages} {245120} (\bibinfo {year} {2007})}\BibitemShut {NoStop}%
\bibitem [{\citenamefont {Holmes-Hewett}\ \emph {et~al.}(2018)\citenamefont {Holmes-Hewett}, \citenamefont {Ullstad}, \citenamefont {Ruck}, \citenamefont {Natali},\ and\ \citenamefont {Trodahl}}]{Holmes2018}%
  \BibitemOpen
  \bibfield  {author} {\bibinfo {author} {\bibfnamefont {W.~F.}\ \bibnamefont {Holmes-Hewett}}, \bibinfo {author} {\bibfnamefont {F.~H.}\ \bibnamefont {Ullstad}}, \bibinfo {author} {\bibfnamefont {B.~J.}\ \bibnamefont {Ruck}}, \bibinfo {author} {\bibfnamefont {F.}~\bibnamefont {Natali}},\ and\ \bibinfo {author} {\bibfnamefont {H.~J.}\ \bibnamefont {Trodahl}},\ }\bibfield  {title} {\bibinfo {title} {Anomalous hall effect in {Sm}n: Influence of orbital magnetism and 4f -band conduction},\ }\bibfield  {journal} {\bibinfo  {journal} {Physical Review B}\ }\textbf {\bibinfo {volume} {98}},\ \href {https://doi.org/10.1103/PhysRevB.98.235201} {10.1103/PhysRevB.98.235201} (\bibinfo {year} {2018})\BibitemShut {NoStop}%
\bibitem [{\citenamefont {Holmes-Hewett}\ \emph {et~al.}(2019)\citenamefont {Holmes-Hewett}, \citenamefont {Buckley}, \citenamefont {Ruck}, \citenamefont {Natali},\ and\ \citenamefont {Trodahl}}]{Holmes2019}%
  \BibitemOpen
  \bibfield  {author} {\bibinfo {author} {\bibfnamefont {W.~F.}\ \bibnamefont {Holmes-Hewett}}, \bibinfo {author} {\bibfnamefont {R.~G.}\ \bibnamefont {Buckley}}, \bibinfo {author} {\bibfnamefont {B.~J.}\ \bibnamefont {Ruck}}, \bibinfo {author} {\bibfnamefont {F.}~\bibnamefont {Natali}},\ and\ \bibinfo {author} {\bibfnamefont {H.~J.}\ \bibnamefont {Trodahl}},\ }\bibfield  {title} {\bibinfo {title} {Optical spectroscopy of {SmN}: Locating the 4f conduction band},\ }\bibfield  {journal} {\bibinfo  {journal} {Physical Review B}\ }\textbf {\bibinfo {volume} {99}},\ \href {https://doi.org/10.1103/PhysRevB.99.205131} {10.1103/PhysRevB.99.205131} (\bibinfo {year} {2019})\BibitemShut {NoStop}%
\bibitem [{\citenamefont {Azeem}(2018)}]{Azeem2018}%
  \BibitemOpen
  \bibfield  {author} {\bibinfo {author} {\bibfnamefont {M.}~\bibnamefont {Azeem}},\ }\bibfield  {title} {\bibinfo {title} {On the optical energy gap of {S}m{N}},\ }\href {https://doi.org/10.1016/j.cjph.2018.07.018} {\bibfield  {journal} {\bibinfo  {journal} {Chinese Journal of Physics}\ }\textbf {\bibinfo {volume} {56}},\ \bibinfo {pages} {1925} (\bibinfo {year} {2018})}\BibitemShut {NoStop}%
\bibitem [{\citenamefont {Azeem}(2019)}]{Azeem2019}%
  \BibitemOpen
  \bibfield  {author} {\bibinfo {author} {\bibfnamefont {M.}~\bibnamefont {Azeem}},\ }\bibfield  {title} {\bibinfo {title} {Quantitative measure of nitrogen vacancy related effects in {S}m{N} and {E}u{N}},\ }\bibfield  {journal} {\bibinfo  {journal} {Advances in Natural Sciences: Nanoscience and Nanotechnology}\ }\textbf {\bibinfo {volume} {10}},\ \href {https://doi.org/10.1088/2043-6254/ab007d} {10.1088/2043-6254/ab007d} (\bibinfo {year} {2019})\BibitemShut {NoStop}%
\bibitem [{\citenamefont {Natali}\ \emph {et~al.}(2012)\citenamefont {Natali}, \citenamefont {Ludbrook}, \citenamefont {Galipaud}, \citenamefont {Plank}, \citenamefont {Granville}, \citenamefont {Preston}, \citenamefont {Do}, \citenamefont {Richter}, \citenamefont {Farrell}, \citenamefont {Reeves}, \citenamefont {Durbin}, \citenamefont {Trodahl},\ and\ \citenamefont {Ruck}}]{Natali2012}%
  \BibitemOpen
  \bibfield  {author} {\bibinfo {author} {\bibfnamefont {F.}~\bibnamefont {Natali}}, \bibinfo {author} {\bibfnamefont {B.}~\bibnamefont {Ludbrook}}, \bibinfo {author} {\bibfnamefont {J.}~\bibnamefont {Galipaud}}, \bibinfo {author} {\bibfnamefont {N.}~\bibnamefont {Plank}}, \bibinfo {author} {\bibfnamefont {S.}~\bibnamefont {Granville}}, \bibinfo {author} {\bibfnamefont {A.}~\bibnamefont {Preston}}, \bibinfo {author} {\bibfnamefont {B.~L.}\ \bibnamefont {Do}}, \bibinfo {author} {\bibfnamefont {J.}~\bibnamefont {Richter}}, \bibinfo {author} {\bibfnamefont {I.}~\bibnamefont {Farrell}}, \bibinfo {author} {\bibfnamefont {R.}~\bibnamefont {Reeves}}, \bibinfo {author} {\bibfnamefont {S.}~\bibnamefont {Durbin}}, \bibinfo {author} {\bibfnamefont {J.}~\bibnamefont {Trodahl}},\ and\ \bibinfo {author} {\bibfnamefont {B.}~\bibnamefont {Ruck}},\ }\bibfield  {title} {\bibinfo {title} {Epitaxial growth and properties of {GdN}, {EuN} and {SmN} thin films},\ }\href {https://doi.org/10.1002/pssc.201100363} {\bibfield  {journal}
  {\bibinfo  {journal} {Physica Status Solidi (C) Current Topics in Solid State Physics}\ }\textbf {\bibinfo {volume} {9}},\ \bibinfo {pages} {605} (\bibinfo {year} {2012})}\BibitemShut {NoStop}%
\bibitem [{\citenamefont {Anton}\ \emph {et~al.}(2013)\citenamefont {Anton}, \citenamefont {Ruck}, \citenamefont {Meyer}, \citenamefont {Natali}, \citenamefont {Warring}, \citenamefont {Wilhelm}, \citenamefont {Rogalev}, \citenamefont {Antonov},\ and\ \citenamefont {Trodahl}}]{Anton2013}%
  \BibitemOpen
  \bibfield  {author} {\bibinfo {author} {\bibfnamefont {E.~M.}\ \bibnamefont {Anton}}, \bibinfo {author} {\bibfnamefont {B.~J.}\ \bibnamefont {Ruck}}, \bibinfo {author} {\bibfnamefont {C.}~\bibnamefont {Meyer}}, \bibinfo {author} {\bibfnamefont {F.}~\bibnamefont {Natali}}, \bibinfo {author} {\bibfnamefont {H.}~\bibnamefont {Warring}}, \bibinfo {author} {\bibfnamefont {F.}~\bibnamefont {Wilhelm}}, \bibinfo {author} {\bibfnamefont {A.}~\bibnamefont {Rogalev}}, \bibinfo {author} {\bibfnamefont {V.~N.}\ \bibnamefont {Antonov}},\ and\ \bibinfo {author} {\bibfnamefont {H.~J.}\ \bibnamefont {Trodahl}},\ }\bibfield  {title} {\bibinfo {title} {Spin/orbit moment imbalance in the near-zero moment ferromagnetic semiconductor {SmN}},\ }\bibfield  {journal} {\bibinfo  {journal} {Physical Review B - Condensed Matter and Materials Physics}\ }\textbf {\bibinfo {volume} {87}},\ \href {https://doi.org/10.1103/PhysRevB.87.134414} {10.1103/PhysRevB.87.134414} (\bibinfo {year} {2013})\BibitemShut {NoStop}%
\bibitem [{\citenamefont {McNulty}\ \emph {et~al.}(2015)\citenamefont {McNulty}, \citenamefont {Anton}, \citenamefont {Ruck}, \citenamefont {Natali}, \citenamefont {Warring}, \citenamefont {Wilhelm}, \citenamefont {Rogalev}, \citenamefont {Soares}, \citenamefont {Brookes},\ and\ \citenamefont {Trodahl}}]{McNulty2015}%
  \BibitemOpen
  \bibfield  {author} {\bibinfo {author} {\bibfnamefont {J.~F.}\ \bibnamefont {McNulty}}, \bibinfo {author} {\bibfnamefont {E.~M.}\ \bibnamefont {Anton}}, \bibinfo {author} {\bibfnamefont {B.~J.}\ \bibnamefont {Ruck}}, \bibinfo {author} {\bibfnamefont {F.}~\bibnamefont {Natali}}, \bibinfo {author} {\bibfnamefont {H.}~\bibnamefont {Warring}}, \bibinfo {author} {\bibfnamefont {F.}~\bibnamefont {Wilhelm}}, \bibinfo {author} {\bibfnamefont {A.}~\bibnamefont {Rogalev}}, \bibinfo {author} {\bibfnamefont {M.~M.}\ \bibnamefont {Soares}}, \bibinfo {author} {\bibfnamefont {N.~B.}\ \bibnamefont {Brookes}},\ and\ \bibinfo {author} {\bibfnamefont {H.~J.}\ \bibnamefont {Trodahl}},\ }\bibfield  {title} {\bibinfo {title} {Twisted phase of the orbital-dominant ferromagnet smn in a {GdN/SmN} heterostructure},\ }\bibfield  {journal} {\bibinfo  {journal} {Physical Review B - Condensed Matter and Materials Physics}\ }\textbf {\bibinfo {volume} {91}},\ \href {https://doi.org/10.1103/PhysRevB.91.174426} {10.1103/PhysRevB.91.174426}
  (\bibinfo {year} {2015})\BibitemShut {NoStop}%
\bibitem [{\citenamefont {McNulty}\ \emph {et~al.}(2021)\citenamefont {McNulty}, \citenamefont {Temst}, \citenamefont {Bael}, \citenamefont {Vantomme},\ and\ \citenamefont {Anton}}]{McNulty2021}%
  \BibitemOpen
  \bibfield  {author} {\bibinfo {author} {\bibfnamefont {J.~F.}\ \bibnamefont {McNulty}}, \bibinfo {author} {\bibfnamefont {K.}~\bibnamefont {Temst}}, \bibinfo {author} {\bibfnamefont {M.~J.~V.}\ \bibnamefont {Bael}}, \bibinfo {author} {\bibfnamefont {A.}~\bibnamefont {Vantomme}},\ and\ \bibinfo {author} {\bibfnamefont {E.~M.}\ \bibnamefont {Anton}},\ }\bibfield  {title} {\bibinfo {title} {Epitaxial growth of (100)-oriented {SmN} directly on (100){S}i substrates},\ }\bibfield  {journal} {\bibinfo  {journal} {Physical Review Materials}\ }\textbf {\bibinfo {volume} {5}},\ \href {https://doi.org/10.1103/PhysRevMaterials.5.113404} {10.1103/PhysRevMaterials.5.113404} (\bibinfo {year} {2021})\BibitemShut {NoStop}%
\bibitem [{\citenamefont {Chan}\ \emph {et~al.}(2016)\citenamefont {Chan}, \citenamefont {Vézian}, \citenamefont {Trodahl}, \citenamefont {Khalfioui}, \citenamefont {Damilano},\ and\ \citenamefont {Natali}}]{Chan2016}%
  \BibitemOpen
  \bibfield  {author} {\bibinfo {author} {\bibfnamefont {J.~R.}\ \bibnamefont {Chan}}, \bibinfo {author} {\bibfnamefont {S.}~\bibnamefont {Vézian}}, \bibinfo {author} {\bibfnamefont {J.}~\bibnamefont {Trodahl}}, \bibinfo {author} {\bibfnamefont {M.~A.}\ \bibnamefont {Khalfioui}}, \bibinfo {author} {\bibfnamefont {B.}~\bibnamefont {Damilano}},\ and\ \bibinfo {author} {\bibfnamefont {F.}~\bibnamefont {Natali}},\ }\bibfield  {title} {\bibinfo {title} {Temperature-induced four-fold-on-six-fold symmetric heteroepitaxy, rocksalt {SmN} on hexagonal {AlN}},\ }\href {https://doi.org/10.1021/acs.cgd.6b01133} {\bibfield  {journal} {\bibinfo  {journal} {Crystal Growth and Design}\ }\textbf {\bibinfo {volume} {16}},\ \bibinfo {pages} {6454} (\bibinfo {year} {2016})}\BibitemShut {NoStop}%
\bibitem [{\citenamefont {Vézian}\ \emph {et~al.}(2016)\citenamefont {Vézian}, \citenamefont {Damilano}, \citenamefont {Natali}, \citenamefont {Khalfioui},\ and\ \citenamefont {Massies}}]{Vezian2016}%
  \BibitemOpen
  \bibfield  {author} {\bibinfo {author} {\bibfnamefont {S.}~\bibnamefont {Vézian}}, \bibinfo {author} {\bibfnamefont {B.}~\bibnamefont {Damilano}}, \bibinfo {author} {\bibfnamefont {F.}~\bibnamefont {Natali}}, \bibinfo {author} {\bibfnamefont {M.~A.}\ \bibnamefont {Khalfioui}},\ and\ \bibinfo {author} {\bibfnamefont {J.}~\bibnamefont {Massies}},\ }\bibfield  {title} {\bibinfo {title} {{AlN} interlayer to improve the epitaxial growth of {SmN} on {GaN} (0001)},\ }\href {https://doi.org/10.1016/j.jcrysgro.2016.06.006} {\bibfield  {journal} {\bibinfo  {journal} {Journal of Crystal Growth}\ }\textbf {\bibinfo {volume} {450}},\ \bibinfo {pages} {22} (\bibinfo {year} {2016})}\BibitemShut {NoStop}%
\bibitem [{\citenamefont {Anton}\ \emph {et~al.}(2023)\citenamefont {Anton}, \citenamefont {Trewick}, \citenamefont {Holmes-Hewett}, \citenamefont {Chan}, \citenamefont {McNulty}, \citenamefont {Butler}, \citenamefont {Ruck},\ and\ \citenamefont {Natali}}]{Anton2023}%
  \BibitemOpen
  \bibfield  {author} {\bibinfo {author} {\bibfnamefont {E.-M.}\ \bibnamefont {Anton}}, \bibinfo {author} {\bibfnamefont {E.}~\bibnamefont {Trewick}}, \bibinfo {author} {\bibfnamefont {W.~F.}\ \bibnamefont {Holmes-Hewett}}, \bibinfo {author} {\bibfnamefont {J.~R.}\ \bibnamefont {Chan}}, \bibinfo {author} {\bibfnamefont {J.~F.}\ \bibnamefont {McNulty}}, \bibinfo {author} {\bibfnamefont {T.}~\bibnamefont {Butler}}, \bibinfo {author} {\bibfnamefont {B.~J.}\ \bibnamefont {Ruck}},\ and\ \bibinfo {author} {\bibfnamefont {F.}~\bibnamefont {Natali}},\ }\bibfield  {title} {\bibinfo {title} {Growth of epitaxial (100)-oriented rare-earth nitrides on (100){LaAlO3}},\ }\bibfield  {journal} {\bibinfo  {journal} {Applied Physics Letters}\ }\textbf {\bibinfo {volume} {123}},\ \href {https://doi.org/10.1063/5.0186522} {10.1063/5.0186522} (\bibinfo {year} {2023})\BibitemShut {NoStop}%
\bibitem [{\citenamefont {Mel{\'e}ndez-Sans}\ \emph {et~al.}(2024)\citenamefont {Mel{\'e}ndez-Sans}, \citenamefont {Pereira}, \citenamefont {Chang}, \citenamefont {Kuo}, \citenamefont {Chen}, \citenamefont {Tjeng},\ and\ \citenamefont {Altendorf}}]{melendez2024influence}%
  \BibitemOpen
  \bibfield  {author} {\bibinfo {author} {\bibfnamefont {A.}~\bibnamefont {Mel{\'e}ndez-Sans}}, \bibinfo {author} {\bibfnamefont {V.}~\bibnamefont {Pereira}}, \bibinfo {author} {\bibfnamefont {C.}~\bibnamefont {Chang}}, \bibinfo {author} {\bibfnamefont {C.-Y.}\ \bibnamefont {Kuo}}, \bibinfo {author} {\bibfnamefont {C.}~\bibnamefont {Chen}}, \bibinfo {author} {\bibfnamefont {L.}~\bibnamefont {Tjeng}},\ and\ \bibinfo {author} {\bibfnamefont {S.}~\bibnamefont {Altendorf}},\ }\bibfield  {title} {\bibinfo {title} {Influence of nitrogen stoichiometry and the role of {S}m 5 d states in {SmN} thin films},\ }\href@noop {} {\bibfield  {journal} {\bibinfo  {journal} {Physical Review B}\ }\textbf {\bibinfo {volume} {110}},\ \bibinfo {pages} {045120} (\bibinfo {year} {2024})}\BibitemShut {NoStop}%
\bibitem [{\citenamefont {Vallejo}\ \emph {et~al.}(2024)\citenamefont {Vallejo}, \citenamefont {Cresswell}, \citenamefont {Buturlim}, \citenamefont {Newell}, \citenamefont {Gofryk},\ and\ \citenamefont {May}}]{vallejo2024synthesis}%
  \BibitemOpen
  \bibfield  {author} {\bibinfo {author} {\bibfnamefont {K.~D.}\ \bibnamefont {Vallejo}}, \bibinfo {author} {\bibfnamefont {Z.~E.}\ \bibnamefont {Cresswell}}, \bibinfo {author} {\bibfnamefont {V.}~\bibnamefont {Buturlim}}, \bibinfo {author} {\bibfnamefont {B.~S.}\ \bibnamefont {Newell}}, \bibinfo {author} {\bibfnamefont {K.}~\bibnamefont {Gofryk}},\ and\ \bibinfo {author} {\bibfnamefont {B.~J.}\ \bibnamefont {May}},\ }\bibfield  {title} {\bibinfo {title} {Synthesis of samarium nitride thin films on magnesium oxide (001) substrates using molecular beam epitaxy},\ }\href@noop {} {\bibfield  {journal} {\bibinfo  {journal} {Crystals}\ }\textbf {\bibinfo {volume} {14}},\ \bibinfo {pages} {765} (\bibinfo {year} {2024})}\BibitemShut {NoStop}%
\bibitem [{\citenamefont {Xie}\ \emph {et~al.}(2025)\citenamefont {Xie}, \citenamefont {Zhang}, \citenamefont {Yin}, \citenamefont {Zheng}, \citenamefont {Ali}, \citenamefont {Younis}, \citenamefont {Ruan},\ and\ \citenamefont {Zeng}}]{xie2025emerging}%
  \BibitemOpen
  \bibfield  {author} {\bibinfo {author} {\bibfnamefont {Y.}~\bibnamefont {Xie}}, \bibinfo {author} {\bibfnamefont {S.-Y.}\ \bibnamefont {Zhang}}, \bibinfo {author} {\bibfnamefont {Y.}~\bibnamefont {Yin}}, \bibinfo {author} {\bibfnamefont {N.}~\bibnamefont {Zheng}}, \bibinfo {author} {\bibfnamefont {A.}~\bibnamefont {Ali}}, \bibinfo {author} {\bibfnamefont {M.}~\bibnamefont {Younis}}, \bibinfo {author} {\bibfnamefont {S.}~\bibnamefont {Ruan}},\ and\ \bibinfo {author} {\bibfnamefont {Y.-J.}\ \bibnamefont {Zeng}},\ }\bibfield  {title} {\bibinfo {title} {Emerging ferromagnetic materials for electrical spin injection: towards semiconductor spintronics},\ }\href@noop {} {\bibfield  {journal} {\bibinfo  {journal} {npj Spintronics}\ }\textbf {\bibinfo {volume} {3}},\ \bibinfo {pages} {10} (\bibinfo {year} {2025})}\BibitemShut {NoStop}%
\bibitem [{\citenamefont {Akinaga}\ and\ \citenamefont {Ohno}(2002)}]{akinaga2002semiconductor}%
  \BibitemOpen
  \bibfield  {author} {\bibinfo {author} {\bibfnamefont {H.}~\bibnamefont {Akinaga}}\ and\ \bibinfo {author} {\bibfnamefont {H.}~\bibnamefont {Ohno}},\ }\bibfield  {title} {\bibinfo {title} {Semiconductor spintronics},\ }\href@noop {} {\bibfield  {journal} {\bibinfo  {journal} {IEEE Transactions on nanotechnology}\ }\textbf {\bibinfo {volume} {1}},\ \bibinfo {pages} {19} (\bibinfo {year} {2002})}\BibitemShut {NoStop}%
\bibitem [{\citenamefont {Dietl}\ \emph {et~al.}(2007)\citenamefont {Dietl}, \citenamefont {Ohno},\ and\ \citenamefont {Matsukura}}]{dietl2007ferromagnetic}%
  \BibitemOpen
  \bibfield  {author} {\bibinfo {author} {\bibfnamefont {T.}~\bibnamefont {Dietl}}, \bibinfo {author} {\bibfnamefont {H.}~\bibnamefont {Ohno}},\ and\ \bibinfo {author} {\bibfnamefont {F.}~\bibnamefont {Matsukura}},\ }\bibfield  {title} {\bibinfo {title} {Ferromagnetic semiconductor heterostructures for spintronics},\ }\href@noop {} {\bibfield  {journal} {\bibinfo  {journal} {IEEE transactions on electron devices}\ }\textbf {\bibinfo {volume} {54}},\ \bibinfo {pages} {945} (\bibinfo {year} {2007})}\BibitemShut {NoStop}%
\bibitem [{\citenamefont {Pot}(2024)}]{pot2024magnetic}%
  \BibitemOpen
  \bibfield  {author} {\bibinfo {author} {\bibfnamefont {C.}~\bibnamefont {Pot}},\ }\emph {\bibinfo {title} {Magnetic Devices Using Rare-Earth Nitrides}},\ \href@noop {} {Ph.D. thesis},\ \bibinfo  {school} {Open Access Te Herenga Waka-Victoria University of Wellington} (\bibinfo {year} {2024})\BibitemShut {NoStop}%
\bibitem [{\citenamefont {Shao}\ and\ \citenamefont {Tsymbal}(2024)}]{shao2024antiferromagnetic}%
  \BibitemOpen
  \bibfield  {author} {\bibinfo {author} {\bibfnamefont {D.-F.}\ \bibnamefont {Shao}}\ and\ \bibinfo {author} {\bibfnamefont {E.~Y.}\ \bibnamefont {Tsymbal}},\ }\bibfield  {title} {\bibinfo {title} {Antiferromagnetic tunnel junctions for spintronics},\ }\href@noop {} {\bibfield  {journal} {\bibinfo  {journal} {npj Spintronics}\ }\textbf {\bibinfo {volume} {2}},\ \bibinfo {pages} {13} (\bibinfo {year} {2024})}\BibitemShut {NoStop}%
\bibitem [{\citenamefont {Felmlee}\ and\ \citenamefont {Eyring}(1968)}]{felmlee1968ternary}%
  \BibitemOpen
  \bibfield  {author} {\bibinfo {author} {\bibfnamefont {T.~L.}\ \bibnamefont {Felmlee}}\ and\ \bibinfo {author} {\bibfnamefont {L.}~\bibnamefont {Eyring}},\ }\bibfield  {title} {\bibinfo {title} {Ternary system samarium-nitrogen-oxygen and the question of lower oxides of samarium},\ }\href@noop {} {\bibfield  {journal} {\bibinfo  {journal} {Inorganic Chemistry}\ }\textbf {\bibinfo {volume} {7}},\ \bibinfo {pages} {660} (\bibinfo {year} {1968})}\BibitemShut {NoStop}%
\bibitem [{\citenamefont {Nordstrom}\ \emph {et~al.}(2023)\citenamefont {Nordstrom}, \citenamefont {Garrett}, \citenamefont {Reddy}, \citenamefont {McElearney}, \citenamefont {Rushing}, \citenamefont {Vallejo}, \citenamefont {Mukherjee}, \citenamefont {Grossklaus}, \citenamefont {Vandervelde},\ and\ \citenamefont {Simmonds}}]{nordstrom2023direct}%
  \BibitemOpen
  \bibfield  {author} {\bibinfo {author} {\bibfnamefont {M.~D.}\ \bibnamefont {Nordstrom}}, \bibinfo {author} {\bibfnamefont {T.~A.}\ \bibnamefont {Garrett}}, \bibinfo {author} {\bibfnamefont {P.}~\bibnamefont {Reddy}}, \bibinfo {author} {\bibfnamefont {J.}~\bibnamefont {McElearney}}, \bibinfo {author} {\bibfnamefont {J.~R.}\ \bibnamefont {Rushing}}, \bibinfo {author} {\bibfnamefont {K.~D.}\ \bibnamefont {Vallejo}}, \bibinfo {author} {\bibfnamefont {K.}~\bibnamefont {Mukherjee}}, \bibinfo {author} {\bibfnamefont {K.~A.}\ \bibnamefont {Grossklaus}}, \bibinfo {author} {\bibfnamefont {T.~E.}\ \bibnamefont {Vandervelde}},\ and\ \bibinfo {author} {\bibfnamefont {P.~J.}\ \bibnamefont {Simmonds}},\ }\bibfield  {title} {\bibinfo {title} {Direct integration of {GaSb} with {GaAs} (111) {A} using interfacial misfit arrays},\ }\href@noop {} {\bibfield  {journal} {\bibinfo  {journal} {Crystal Growth \& Design}\ }\textbf {\bibinfo {volume} {23}},\ \bibinfo {pages} {8670} (\bibinfo {year} {2023})}\BibitemShut {NoStop}%
\bibitem [{\citenamefont {Zhang}\ \emph {et~al.}(2024)\citenamefont {Zhang}, \citenamefont {Hilse}, \citenamefont {Auker}, \citenamefont {Gray},\ and\ \citenamefont {Law}}]{zhang2024growth}%
  \BibitemOpen
  \bibfield  {author} {\bibinfo {author} {\bibfnamefont {Q.}~\bibnamefont {Zhang}}, \bibinfo {author} {\bibfnamefont {M.}~\bibnamefont {Hilse}}, \bibinfo {author} {\bibfnamefont {W.}~\bibnamefont {Auker}}, \bibinfo {author} {\bibfnamefont {J.}~\bibnamefont {Gray}},\ and\ \bibinfo {author} {\bibfnamefont {S.}~\bibnamefont {Law}},\ }\bibfield  {title} {\bibinfo {title} {Growth conditions and interfacial misfit array in snte (111) films grown on inp (111) a substrates by molecular beam epitaxy},\ }\href@noop {} {\bibfield  {journal} {\bibinfo  {journal} {ACS Applied Materials \& Interfaces}\ }\textbf {\bibinfo {volume} {16}},\ \bibinfo {pages} {48598} (\bibinfo {year} {2024})}\BibitemShut {NoStop}%
\bibitem [{\citenamefont {Meyer}\ \emph {et~al.}(2008)\citenamefont {Meyer}, \citenamefont {Ruck}, \citenamefont {Zhong}, \citenamefont {Granville}, \citenamefont {Preston}, \citenamefont {Williams},\ and\ \citenamefont {Trodahl}}]{Meyer2008}%
  \BibitemOpen
  \bibfield  {author} {\bibinfo {author} {\bibfnamefont {C.}~\bibnamefont {Meyer}}, \bibinfo {author} {\bibfnamefont {B.~J.}\ \bibnamefont {Ruck}}, \bibinfo {author} {\bibfnamefont {J.}~\bibnamefont {Zhong}}, \bibinfo {author} {\bibfnamefont {S.}~\bibnamefont {Granville}}, \bibinfo {author} {\bibfnamefont {A.~R.}\ \bibnamefont {Preston}}, \bibinfo {author} {\bibfnamefont {G.~V.}\ \bibnamefont {Williams}},\ and\ \bibinfo {author} {\bibfnamefont {H.~J.}\ \bibnamefont {Trodahl}},\ }\bibfield  {title} {\bibinfo {title} {Near-zero-moment ferromagnetism in the semiconductor smn},\ }\bibfield  {journal} {\bibinfo  {journal} {Physical Review B - Condensed Matter and Materials Physics}\ }\textbf {\bibinfo {volume} {78}},\ \href {https://doi.org/10.1103/PhysRevB.78.174406} {10.1103/PhysRevB.78.174406} (\bibinfo {year} {2008})\BibitemShut {NoStop}%
\bibitem [{\citenamefont {Holmes-Hewett}\ \emph {et~al.}(2023)\citenamefont {Holmes-Hewett}, \citenamefont {Van~Koughnet}, \citenamefont {Miller}, \citenamefont {Trewick}, \citenamefont {Ruck}, \citenamefont {Trodahl},\ and\ \citenamefont {Buckley}}]{Holmes2023}%
  \BibitemOpen
  \bibfield  {author} {\bibinfo {author} {\bibfnamefont {W.}~\bibnamefont {Holmes-Hewett}}, \bibinfo {author} {\bibfnamefont {K.}~\bibnamefont {Van~Koughnet}}, \bibinfo {author} {\bibfnamefont {J.}~\bibnamefont {Miller}}, \bibinfo {author} {\bibfnamefont {E.}~\bibnamefont {Trewick}}, \bibinfo {author} {\bibfnamefont {B.}~\bibnamefont {Ruck}}, \bibinfo {author} {\bibfnamefont {H.}~\bibnamefont {Trodahl}},\ and\ \bibinfo {author} {\bibfnamefont {R.}~\bibnamefont {Buckley}},\ }\bibfield  {title} {\bibinfo {title} {Indications of a ferromagnetic quantum critical point in smn 1-$\delta$},\ }\href@noop {} {\bibfield  {journal} {\bibinfo  {journal} {Scientific Reports}\ }\textbf {\bibinfo {volume} {13}},\ \bibinfo {pages} {19775} (\bibinfo {year} {2023})}\BibitemShut {NoStop}%
\bibitem [{\citenamefont {Karpov}\ \emph {et~al.}(2000)\citenamefont {Karpov}, \citenamefont {Talalaev}, \citenamefont {Makarov}, \citenamefont {Grandjean}, \citenamefont {Massies},\ and\ \citenamefont {Damilano}}]{karpov2000surface}%
  \BibitemOpen
  \bibfield  {author} {\bibinfo {author} {\bibfnamefont {S.~Y.}\ \bibnamefont {Karpov}}, \bibinfo {author} {\bibfnamefont {R.}~\bibnamefont {Talalaev}}, \bibinfo {author} {\bibfnamefont {Y.~N.}\ \bibnamefont {Makarov}}, \bibinfo {author} {\bibfnamefont {N.}~\bibnamefont {Grandjean}}, \bibinfo {author} {\bibfnamefont {J.}~\bibnamefont {Massies}},\ and\ \bibinfo {author} {\bibfnamefont {B.}~\bibnamefont {Damilano}},\ }\bibfield  {title} {\bibinfo {title} {Surface kinetics of gan evaporation and growth by molecular-beam epitaxy},\ }\href@noop {} {\bibfield  {journal} {\bibinfo  {journal} {Surface science}\ }\textbf {\bibinfo {volume} {450}},\ \bibinfo {pages} {191} (\bibinfo {year} {2000})}\BibitemShut {NoStop}%
\bibitem [{\citenamefont {Mart{\'\i}n-Palma}\ and\ \citenamefont {Mart{\'\i}nez-Duart}(2017)}]{martin2017nanotechnology}%
  \BibitemOpen
  \bibfield  {author} {\bibinfo {author} {\bibfnamefont {R.~J.}\ \bibnamefont {Mart{\'\i}n-Palma}}\ and\ \bibinfo {author} {\bibfnamefont {J.}~\bibnamefont {Mart{\'\i}nez-Duart}},\ }\href@noop {} {\emph {\bibinfo {title} {Nanotechnology for microelectronics and photonics}}}\ (\bibinfo  {publisher} {Elsevier},\ \bibinfo {year} {2017})\BibitemShut {NoStop}%
\bibitem [{\citenamefont {Gupta}\ \emph {et~al.}(2012)\citenamefont {Gupta}, \citenamefont {Suresh},\ and\ \citenamefont {Nigam}}]{gupta2012magnetocaloric}%
  \BibitemOpen
  \bibfield  {author} {\bibinfo {author} {\bibfnamefont {S.~B.}\ \bibnamefont {Gupta}}, \bibinfo {author} {\bibfnamefont {K.}~\bibnamefont {Suresh}},\ and\ \bibinfo {author} {\bibfnamefont {A.}~\bibnamefont {Nigam}},\ }\bibfield  {title} {\bibinfo {title} {Magnetocaloric and magnetotransport properties in {RRhSn} ({R}= {Tb-Tm}) series},\ }\href@noop {} {\bibfield  {journal} {\bibinfo  {journal} {Journal of Applied Physics}\ }\textbf {\bibinfo {volume} {112}},\ \bibinfo {pages} {103909} (\bibinfo {year} {2012})}\BibitemShut {NoStop}%
\end{thebibliography}%

\end{document}